\documentclass{iopart}
\usepackage{amsmath,amssymb,graphicx,bbm}
\def\id{\mathbbm I}
\newcommand{\smb}{{\scriptscriptstyle B}}
\newcommand{\sma}{{\scriptscriptstyle A}}
\newcommand{\smbo}{{\scriptscriptstyle B1}}
\newcommand{\smao}{{\scriptscriptstyle A1}}
\newcommand{\smbt}{{\scriptscriptstyle B2}}
\newcommand{\smat}{{\scriptscriptstyle A2}}
\begin{document}
\title{Two quantum Simpson's paradoxes}
\author{Matteo G. A. Paris}
\address{Dipartimento di Fisica dell'Universit\`a degli Studi di
Milano, I-20133 Milano, Italy}
\date{\tt notes  -- mid August 2011}
\date{\tt notes  -- end September 2011}
\date{\tt notes  -- end December 2011}
\begin{abstract}
The so-called  Simpson's "paradox", or Yule-Simpson (YS) effect, occurs
in classical statistics when the correlations that are present among
different sets of samples are reversed if the sets are combined
together, thus ignoring one or more {\em lurking} variables. Here we
illustrate the occurrence of two analogue effects in quantum
measurements. The first, which we term {\em quantum-classical} YS effect, 
may occur with quantum limited measurements and with lurking
variables coming from the mixing of states, whereas the second, here
referred to as {\em quantum-quantum} YS effect, may take place
when coherent superpositions of quantum states are allowed.  By analyzing
quantum measurements on low dimensional systems (qubits and qutrits), we
show that the two effects may occur independently, and that the
quantum-quantum YS effect is more likely to occur than the corresponding 
quantum-classical one. We also found that there exist classes of 
superposition states for which the quantum-classical YS effect cannot 
occur for any measurement and, at the same time, the quantum-quantum YS  
effect takes place in a consistent fraction of the possible
measurement settings. The occurrence of the effect in the presence of 
partial coherence is discussed as well as its possible implications
for quantum hypothesis testing.
\end{abstract}
\section{Introduction}
In classical statistics, the so-called Simpson's "paradox", also
referred to as the Yule-Simpson effect \cite{Sim51,Bly72,Bic75}, occurs when
the correlations observed in different groups are reversed when the
groups are combined together.  The typical examples come from social- or
medical-science \cite{Mes81,YSs1,YSm1,YSm2}: Suppose you are given two
samples, $A$ and $B$, on which the rates of success of two events, say,
the success of two different therapies or of two ways of applying for a
job, are given by $p_\sma$ and $q_\smb$ respectively, with $p_\sma <
q_\smb$. The rates of success express the degree of correlations between
the events under consideration and the characteristic features of the
samples $A$ and $B$. Then, it may occur that by splitting the initial
samples in groups, say $A_1,A_2$ and $B_1, B_2$, according to the the
value of a certain {\em lurking variable} (say, gender, age,
geographical localization,..) the ordering of the rates of success, and
thus of the correlations, is reversed, in formula $p_\smao > q_\smbo$
and $p_\smat > q_\smbt$. As for example, a certain therapy may appear
{\em good for women} and {\em good for men}, but {\em bad for people}
\cite{YSm2}.
\par
Actually, there is no mathematical paradox: The YS effect
arises from a hidden correlation, and it may be
explained in terms of the relative weights of the groups $A_1,A_2$ and
$B_1, B_2$, due to different sizes of the samples in the groups $A_j$ and
$B_j$ \cite{Lip06,Her11}. An example illustrating the point is reported
in the appendix. 
On the other hand, the practical consequences of the effect are
indeed counterintuitive for decision making, since the aggregated data
and the partitioned ones are, in fact, suggesting opposite strategies
\cite{Ald95,Cox03}. Other examples arises in
game theory, where it may be shown that 
two losing strategies may be randomly combined to form a winning one 
\cite{Par00,Buc02,Kay03,Ast05} realizing, in fact, a variation of the 
YS effect.
From the operational point of view there is an
additional reason for the emergence of the YS effect, which may
summarized by saying the two events are incompatible: once a therapy has
been administered, or an application has been submitted, there is no way
to determine the outcome of the other options for the same individual
\cite{Gud88}.  In other words, there is no way to assign a definite
meaning to the rate of success of the two events simultaneously. Loosely
speaking, this fact suggests that other incompatibilities \cite{Coh86}, as those
arising from the quantum mechanical description of measurements, may
play a role for the occurrence of the YS effect.
\par
In this communication, we address this possibility in details and investigate
the occurrence of the YS effect in systems subjected to the laws of
quantum mechanics. In particular, we analyze the role of mixing and
superposition of quantum states, and illustrate the occurrence of two
kinds of YS effects in quantum measurements. The first, which we term
{\em quantum-classical} YS effect (QCYS), may occur with quantum limited
measurements and mixing of states, whereas the second, here referred to
as {\em quantum-quantum} YS effect (QQYS), may take place when
superpositions of quantum states are allowed. We describe both effects
in some details, prove that they may occur independently, and present 
a class of states for which QQYS effect occurs when, for
the same values of the involved parameters, the QCYS one  does not.
\section{The quantum-classical YS effect}
Let us consider two quantum tests, i.e. two binary probability 
operator-valued measures (POVMs) $A=\{\Pi_\sma,
\id-\Pi_\sma\}$ and $B=\{\Pi_\smb, \id-\Pi_\smb\}$ aimed at describing
the occurrence of certain events $A$ and $B$. Given a quantum state
$|\psi\rangle$, the expectation value $\langle \psi| \Pi_j|\psi\rangle$
returns the probability of the event $j=A, B$ on the state
$|\psi\rangle$, and thus represents the correlations between the
occurrence of the event and the preparation of the system.
\par
We assume that the system under investigation may be prepared 
in two possible states $|\psi_j\rangle$, $j=1,2$ and that the
event $A$ is more likely to happen than the event $B$ 
for both preparations, i.e.
$$
p_j=\langle \psi_j| \Pi_\sma|\psi_j\rangle
> q_j=\langle \psi_j| \Pi_\smb|\psi_j\rangle\,.
$$
We have the Yule-Simpson effect whenever it occurs that for some 
choice of the mixing parameters (which play the role of relative 
weights due to different sample sizes) 
the event $B$ become more probable 
for the system prepared in mixed state
$\varrho_\beta = \cos^2\beta |\psi_1\rangle\langle \psi_1| + \sin^2\beta 
|\psi_2\rangle\langle \psi_2|$
rather than for the system prepared in mixed state
$\varrho_\alpha = \cos^2\alpha |\psi_1\rangle\langle \psi_1| + \sin^2\alpha 
|\psi_2\rangle\langle \psi_2|$
i.e. that  
\begin{align}
p \equiv\hbox{Tr}[\varrho_\alpha\,\Pi_\sma] < q \equiv\hbox{Tr}[\varrho_\beta\,\Pi_\smb]
\end{align}
where
$
\hbox{Tr}[\varrho_\alpha\,\Pi_\sma] = \cos^2\alpha\, p_1 +
\sin^2\alpha\, p_2$ and 
$ \hbox{Tr}[\varrho_\beta\,\Pi_\smb] = \cos^2\beta\, q_1 +
\sin^2\beta\, q_2$ (see Table 1 for a summary). 
The condition $p<q$ may be written as
\begin{align}
\cos^2\beta\, (q_1-q_2) >  \cos^2\alpha\, (p_1-p_2) +  (p_2 -q_2)\,.
\label{cyscond}
\end{align}
and it is satisfied by 
$\cos^2\beta \gtrless T$ if $p_1-p_2 \gtrless 0$ and 
$q_1-q_2\gtrless0$ \cite{Had97},  
where $T= a\,\cos^2\alpha +  b$, $a= (p_1-p_2)/(q_1-q_2)$ and 
$b=(p_2-q_2)/(q_1-q_2)$.
We refer to this form of the Yule-Simpson effect
as to {\em quantum-classical} Simpson's paradox since it happens
in quantum-limited measurements, however with the lurking
variable coming from the classical mixing of two quantum states.
\begin{table}[h!]
\begin{tabular}{cc|ccc|cc}
  & &$\Pi_\sma$& & $\Pi_\smb$&  & \\
 \hline
$\cos^2\alpha$ & $|\psi_1\rangle$ &$p_1$ &$>$& $q_1$ & $|\psi_1\rangle$
& $\cos^2\beta$ \\
$\sin^2\alpha$ & $|\psi_2\rangle$& $p_2$ &$>$& $q_2$ & $|\psi_2\rangle$
& $\sin^2\beta$ \\
\hline
 & $\varrho_\alpha$ & $p$ &$<$ & $q$ & $\varrho_\beta$ &  
\end{tabular}
\caption{The quantum-classical Yule-Simpson effect}
\end{table}
\par
As mentioned in the introduction, unequal mixing of the two states 
is needed for the occurrence of the YS effect. Indeed, by putting 
$\cos^2\alpha=\cos^2\beta=\frac12$ in Eq. (\ref{cyscond}), one
obtains $\varrho_\alpha=\varrho_\beta$, and thus the condition 
$\frac12 (p_1+p_2)<\frac12 (q_1+q_2)$,
which is never satisfied within the initial assumptions $p_j>q_j$.
\subsection{The QCYS effect in quantum hypothesis testing}
In this section we provide an example, illustrating the realization of the 
QCYS effect in a decision problem involving quantum measurements and 
a finite number of runs.
\par
Suppose that you are given a black box, which may implement two possible 
dichotomic measurements $A=\{\Pi_\sma,
\id-\Pi_\sma\}$ and $B=\{\Pi_\smb, \id-\Pi_\smb\}$ 
on a given system, and you have to infer 
which  measurement has been performed on the basis of the results of the 
measurement. To this aim, you may  {\em probe} the measuring box 
$M$ times, and in each run you have at disposal two possible preparations 
of the system, say $|\psi_j\rangle$, $j=1,2$. 
In order to have a specific example in mind we may consider the
case of a Sten-Gerlach apparatus, which may realize the measurement of a
spin component along a given direction $\theta$, i.e. 
$\Pi_\sma=|0\rangle_\theta{}_\theta\langle0|$, 
or along a slightly tilted one $\theta'$, i.e. 
$\Pi_\smb=|0\rangle_{\theta'}{}_{\theta'}\langle0|$.
We have access to a pair of possible preparations 
of the spin system, and we have to infer which component has been
actually measured on the basis of the number of, say, upper spots
recorded after $M=M_1+M_2$ repeated measurements, where $M_j$ is 
the number of runs where the system was prepared in the state
$|\psi_j\rangle$. 
\par
If we know which preparation $|\psi_j\rangle$ is used in each run then,
using the notation of the previous section and assuming $p_j> q_j$, 
we would always infer that the box is performing measurement $A$,
independently on the values of $M_1$ and $M_2$.
On the other hand, if we ignore the information about which state has
been sent to the box in each run, i.e. we aggregate spots, then we may
reach the opposite conclusion, depending on the relative weight
$M_1/M_2$ of the samples. More explicitly, we have the QCYS effect
whenever using $M_1$ or $M_1^\prime$ times the probe state $|\psi_1\rangle$,
the quantities $\cos^2\alpha=M_1/M$ and $\cos^2\beta=M_1^\prime/M$ satisfy
Eq. (\ref{cyscond}). As in the classical case there is no
mathematical paradox: still the aggregated data and the 
partitioned ones may, in fact, suggest opposite conclusions.
\section{ The quantum-quantum YS effect} 
Here we address situations where the lurking 
variables are coming from the coherent superposition of 
quantum states, rather than from their mixing, and discuss
the occurrence of the corresponding
{\em quantum-quantum} Simpson's paradox. 
\par
Let us consider a situation where the system under investigation, 
besides the states $|\psi_j\rangle$, $j=1,2$, may be prepared in 
any superposition of the form
\begin{align}\label{superp}
|\psi_\gamma\rangle &= \frac{1}{\sqrt{N_\gamma}}\big[\cos\gamma\, 
|\psi_1\rangle+ e^{- i \phi_\gamma}\sin\gamma\, |\psi_2\rangle\big]
\end{align}
where $N_\gamma = 1 +  \kappa_\psi \cos
(\phi_\gamma-\varphi_\kappa)\sin 2\gamma$ is the normalization,  and 
$\kappa_\psi\,e^{i\varphi_\kappa} = \langle\psi_1|\psi_2\rangle$
is the overlap between the two initial preparations.
We assume, as in the previous case, that $p_j>q_j$ $j=1,2$. 
Then, the YS effect occurs whenever, for two different superpositions
$|\psi_\alpha\rangle$ and $|\psi_\beta\rangle$, it happens that 
\begin{align}
\langle\psi_\alpha|\Pi_\sma|\psi_\alpha\rangle\equiv P
<Q\equiv\langle\psi_\beta|\Pi_\smb|\psi_\beta\rangle\,,
\label{PQ1}
\end{align}
where 
\begin{align}
P&= \frac{1}{N_\alpha}\left[\cos^2\alpha\,p_1 + \sin^2\alpha\, p_2 
+ \kappa_\sma \cos(\phi_\alpha - \varphi_\sma)\sin 2\alpha\right] \notag \\
&= \frac{p+\kappa_\sma \cos(\phi_\alpha - \varphi_\sma)\sin 2\alpha}
{1 +  \kappa_\psi \cos (\phi_\alpha-\varphi_\kappa)\sin 2\alpha} \notag \\
Q&= \frac{1}{N_\beta} \left[\cos^2\beta\,q_1 + \sin^2\beta\, q_2 
+ \kappa_\smb \cos(\phi_\beta - \varphi_\smb)\sin 2\beta\right] \notag \\
&= \frac{q+\kappa_\smb \cos(\phi_\beta - \varphi_\smb)\sin 2\beta}
{1 +  \kappa_\psi \cos (\phi_\beta-\varphi_\kappa)\sin 2\beta} 
\label{PQ2}
\end{align}
with
$
\langle\psi_1|\Pi_j|\psi_2\rangle = \kappa_j e^{-i
\varphi_j} \quad j={A}, {B}$ (see Table 2 for a summary).
In the following we will investigate whether the two effects
may occur independently (i.e. $p<q$ when $P>Q$, and viceversa)
and, in particular, whether or not the quantum-quantum Yule-Simpson effect 
may occur when, for the same values of the involved parameters, 
the quantum-classical effect does not.
\begin{table}[h!]
\begin{tabular}{cc|ccc|cc}
  & &$\Pi_\sma$& & $\Pi_\smb$&  & \\
 \hline
$\cos\alpha$ & $|\psi_1\rangle$ &$p_1$ &$>$& $q_1$ & $|\psi_1\rangle$
& $\cos\beta$ \\
$\sin\alpha\,e^{i\phi_\alpha}$ & $|\psi_2\rangle$& $p_2$ &$>$& $q_2$ & $|\psi_2\rangle$
& $\sin\beta\,e^{i\phi_\beta}$ \\
\hline
 & $|\psi_\alpha\rangle$ & $P$ &$<$ & $Q$ & $|\psi_\beta\rangle$ &  
\end{tabular}
\caption{The quantum-quantum Yule-Simpson effect}
\end{table}
\section{Results for low dimensional systems}
In order to gain a quantitative insight into both effects let 
us start by focusing on bidimensional systems (qubits). 
In this case we may write the initial states as 
$|\psi_j\rangle = \cos\theta_j |0\rangle + e^{i \phi_j} \sin\theta_j
|1\rangle$, $j=1,2$ where $\{|0\rangle,|1\rangle\}$ is a basis for the 
Hilbert space, and the POVMs as $\Pi_j = \frac12 (a_j + {\boldsymbol
r}_j\cdot {\boldsymbol \sigma})$, $j={A,B}$, where 
${\boldsymbol r}=\{r_1, r_2, r_3\}$, and
${\boldsymbol \sigma}=\{\sigma_1, \sigma_2, \sigma_3\}$ are Pauli
matrices.
The constraint of positivity for the $\Pi_j$'s is expressed by 
the conditions $|{\boldsymbol r_j}| \leq a_j \leq 2-|{\boldsymbol
r_j}|$. Projective measurements are those individuated by the 
relations $|{\boldsymbol r_j}| =a_j=1$.
\par
Before going to the general case, let us consider the specific example
where $A$ and $B$ are mutually exclusive events. Classically, there is
no YS effect in this case, since no lurking variable may change the
correlations. Quantum mechanically, the two mutually exclusive 
events are described by projective measurements on two 
orthogonal states. For qubits they span
the entire Hilbert space, i.e. $\Pi_\sma+\Pi_\smb=\id$. 
As a  consequence, we have $q_j=1-p_j$ and, in turn, $p>q$ if $p_j>q_j$, 
i.e no QCYS effect.
On the other hand, by looking at Eqs. (\ref{PQ1}) and (\ref{PQ2}), 
one sees that QQYS effect takes place whenever we have $P<\frac12$ 
for $p>\frac12$, e.g. upon choosing $\Pi_\sma=|0\rangle\langle 0|$ and 
$\Pi_\smb=|1\rangle\langle 1|$, for $\theta_1=0$, $\theta_2=\pi/4$,
$\phi_1=\phi_2=0$, $\alpha=\pi/2$, $\beta=\pi/4$, and 
$\phi_\alpha=\phi_\beta=\pi$. 
\par
\begin{figure}[ht!]
\includegraphics[width=0.44\columnwidth]{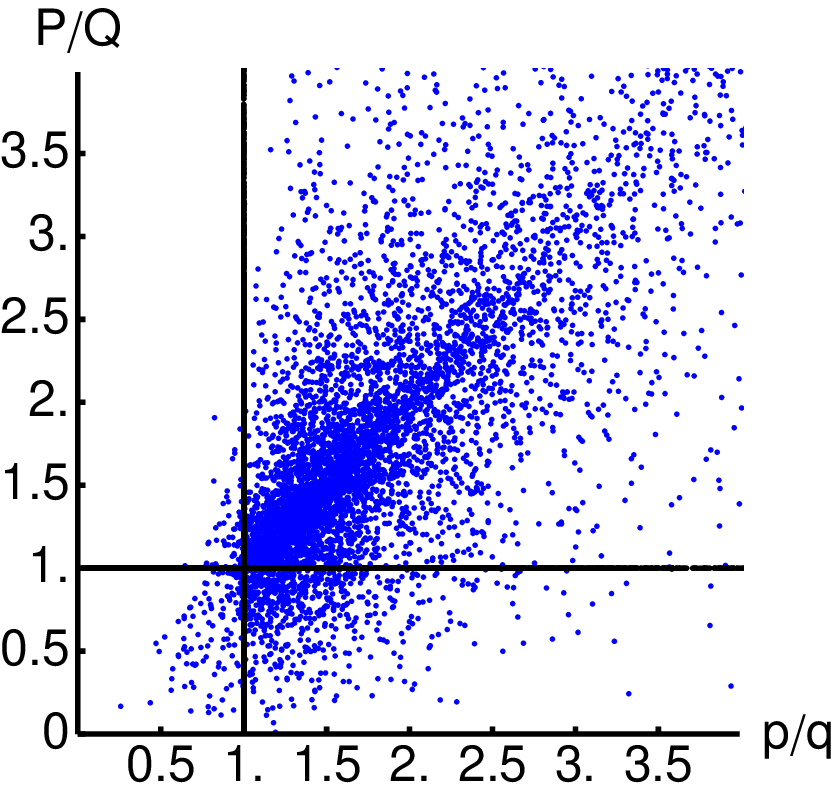}
\includegraphics[width=0.44\columnwidth]{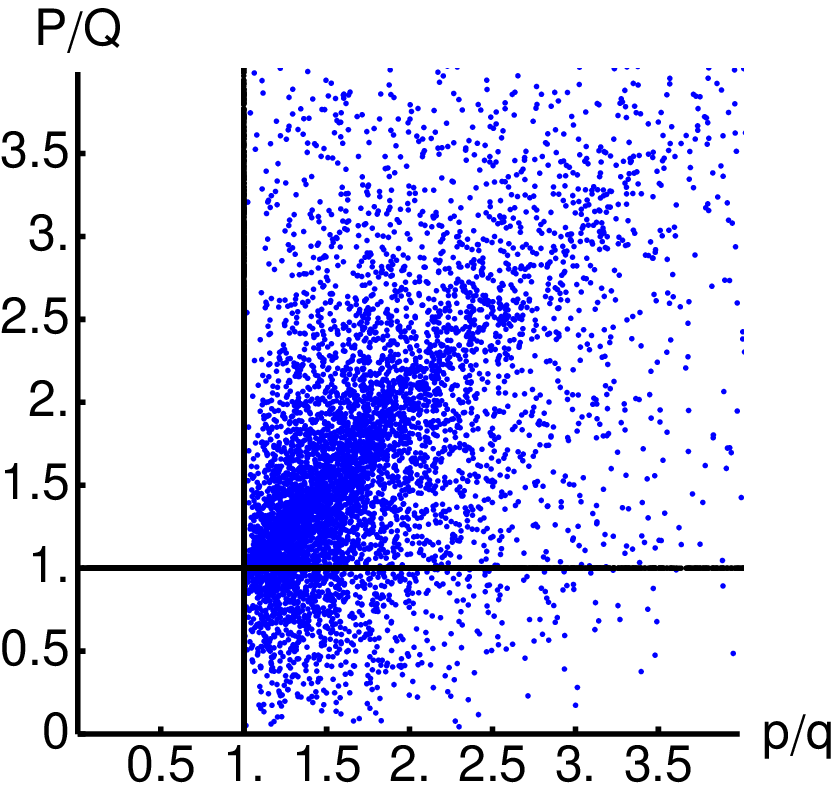}
\caption{(Color online) Quantum-quantum YS effect vs 
quantum-classical YS effect for 
qubit systems. Left: the 
ratio $P/Q$ as a function of $p/q$ for $10^4$ randomly 
generated sets of states, measurements, and lurking variables. 
Right: the same plot for fixed values of the mixing parameters 
$\alpha=\beta=\pi/4$.}\label{f:d2}
\end{figure}

More generally, the occurrence of QCYS and QQYS may be  
investigated in terms of the four + eight + four parameters 
describing the initial states
$|\psi_j\rangle$, $j=1,2$, the POVMs $\Pi_j$, $j={A, B}$,  
and the lurking variables
$\alpha, \beta, \phi_\alpha, \phi_\beta$.
In Fig. \ref{f:d2} we report results for random states and measurements,
generated to satisfy the constraints $p_j>q_j$, $j=1,2$, and for random
lurking variables.  In particular, we show the ratio $P/Q$ as a function
of the ratio $p/q$.  The square region $p/q \in (0,1) \times P/Q \in
(0,1)$ corresponds to the occurrence of both the QC and QQ YS effects,
whereas the rectangular regions $ (0,1)\times [1, \infty)$ and $
[1,\infty)\times (0,1)$ correspond to cases when solely the QC effect, 
or the QQ one, takes place, respectively. Finally, the 
region $[1,\infty)\times [1,\infty)$
corresponds to situations when we have no YS effects. On the left, we
show the results for random states and measurements, and for 
random lurking variables
$\alpha,\beta\in[0,\pi/2]$, $\phi_\alpha,\phi_\beta\in[0, 2\pi]$,
whereas on the right we consider only situations for which
$\alpha=\beta=\pi/4$. 
\par
As it is apparent from the left plot, the two effects may occur
independently, i.e. we may have $p<q$ when $P>Q$, and viceversa.
For random settings, i.e. random states, measurements and lurking variables,
the QQYS effect is more likely to occur than the corresponding 
CCYS effect.  Upon generating a sample of $10^4$ random settings, we found  
QCYS effect without the QQYS one in about $0.97\%$ of cases, just the
reverse in about $10.9\%$ of cases, and both effects 
in about $1.7\%$ of cases.
The right plot refers to random initial states and measurements, and 
to lurking variables with equal mixing parameters $\cos^2\alpha
=\cos^2\beta=\frac12$. In this case, as mentioned before, 
the QCYS effect cannot occur for any measurement. On the other hand, 
upon considering quantum superpositions, we have 
that the QQYS effect occurs for a consistent fraction of the 
possible settings (about $14\%$). If we consider only 
projective measurements, the rates of occurrence of both the YS effects
increases, but the overall picture is not qualitatively modified.
Similar results are also obtained by considering three-dimensional 
quantum systems (qutrits).
\subsection{Partial coherence and the generalized QQYS effect}
Let us now assume that besides the complete mixture $\varrho_\gamma$, and
the coherent superposition $|\psi_\gamma\rangle$, the system may be also 
prepared in partially coherent superpositions of the two initial states 
$|\psi_j\rangle$, i.e. we consider the general class of states 
$\varrho_{\lambda\gamma} 
= \lambda
|\psi_\gamma\rangle\langle\psi_\gamma | + (1-\lambda) \varrho_\gamma$, 
where $\gamma=\alpha, \beta$, $0\leq \lambda \leq1$. The family
$\varrho_{\lambda\gamma}$ continuously connects the coherent 
superposition to the complete mixture, and we now address the occurrence
of the corresponding generalized QQYS effect, which takes place, 
assuming again $p_j>q_j$, $j=1,2$, when we have $P_\lambda<Q_\lambda$, where
$P_\lambda = \hbox{Tr}\left[\varrho_{\lambda\alpha}\,\Pi_\sma \right]
= \lambda P + (1-\lambda) p$, $Q_\lambda = 
\hbox{Tr}\left[\varrho_{\lambda\beta}\,\Pi_\smb \right]
= \lambda Q + (1-\lambda) q$. Using these expressions 
and introducing the threshold value 
$$ \lambda_{\rm th} = 
\left(1-\frac{P-Q}{p-q}\right)^{-1}\,,$$
one easily see that the QQYS effect persists in the range
$\lambda \in (\lambda_{\rm th},1)$ if $p-q>P-Q$,
or $\lambda \in (0,\lambda_{\rm th})$ if $p-q<P-Q$.
\begin{figure}[ht!]
\includegraphics[width=0.31\columnwidth]{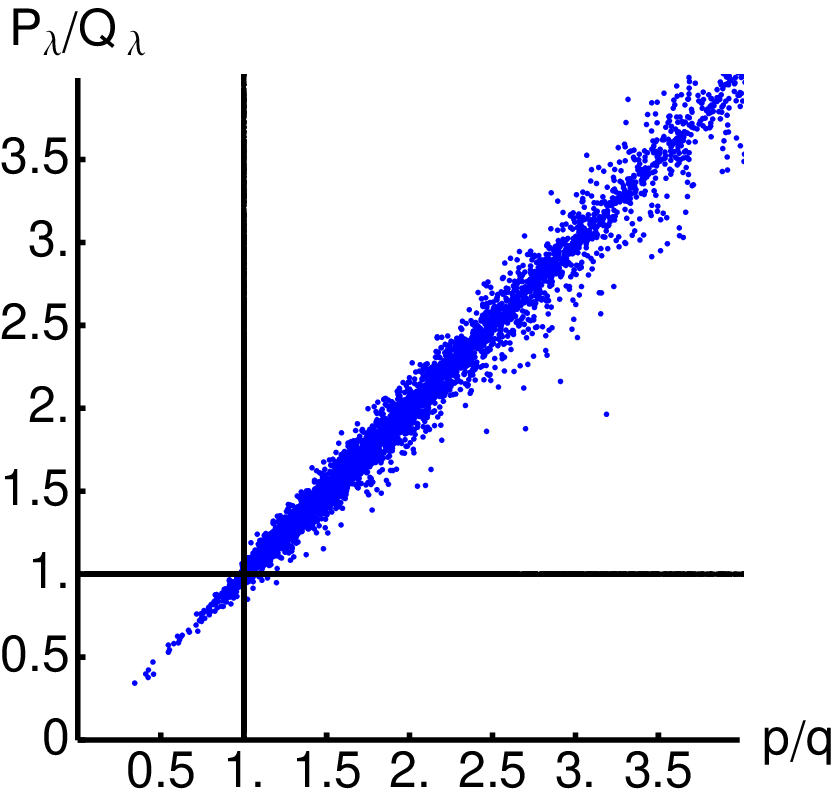}
\includegraphics[width=0.31\columnwidth]{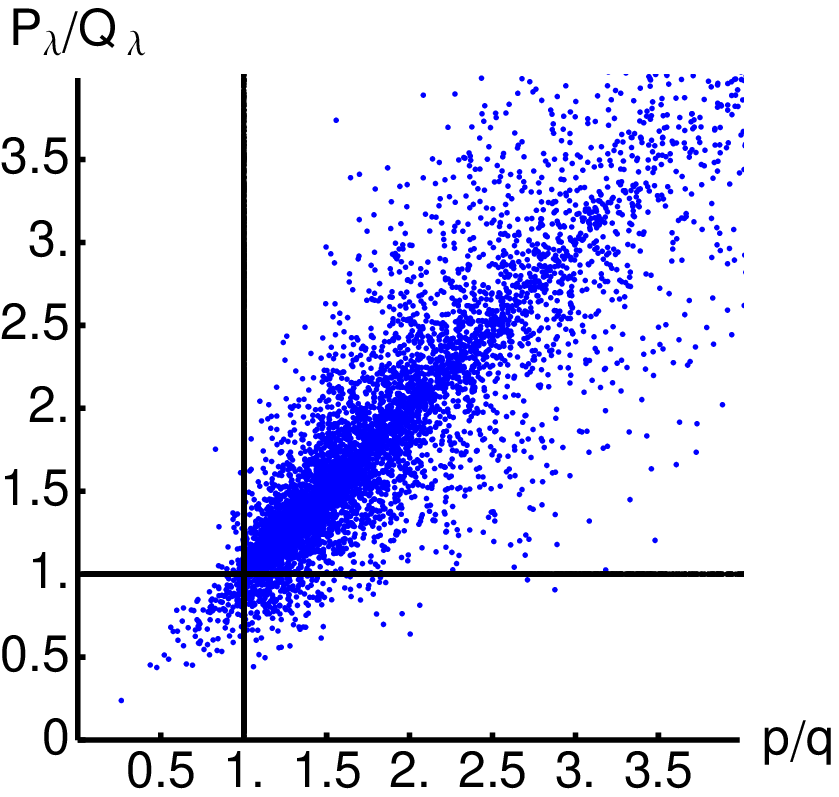}
\includegraphics[width=0.31\columnwidth]{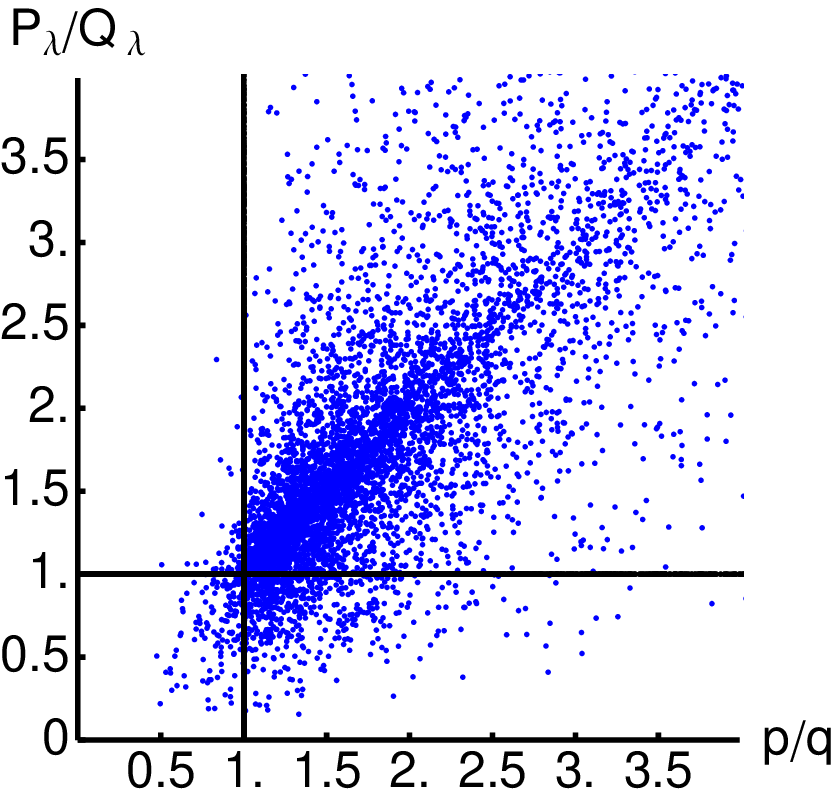}
\caption{(Color online) Generalized quantum-quantum YS 
effect vs quantum-classical YS effect for 
qubit systems. The ratio $P_\lambda/Q_\lambda$ is shown as 
a function of $p/q$ for $10^4$ randomly generated sets of 
states, measurements, and lurking variables. From left to right:
$\lambda=0.1, 0.5, 0.9$ respectively.}\label{f:lam}
\end{figure}
\par
The transition from the QC to the QQ effect is illustrated in Fig.
\ref{f:lam}, where we show the ratio $P_\lambda/Q_\lambda$ as a function
of the ratio $p/q$ for $10^4$ sets of randomly generated states, measurements and
lurking variables (as in Fig. \ref{f:d2}), and for different values 
of $\lambda$.
\section{Conclusions}
The Yule-Simpson is a paradigmatic paradoxical effect occurring 
in data aggregation in complex systems, and found applications 
in the physical description of, e.g. social processes \cite{Gal90}, 
complex dynamics \cite{Alm05}, and game theory \cite{mcd09}. 
In this communication, we have illustrated the occurrence of 
the YS effect in quantum measurements, where the lurking variables 
are coming either from the mixing or the superposition of quantum 
states. By analyzing low dimensional systems, we have found
that the two effects may occur independently, and that the
quantum-quantum YS effect is more likely to occur than the corresponding 
quantum-classical one for the same set of states and measurements. 
We have also discussed an example, illustrating the occurence of
the quantum-classical YS effect in a decision problem involving quantum
measurements.
\appendix
\section{An example of YS effect: sex bias in graduate admissions}
One of the best known real life examples of Simpson's paradox occurred
when the University of California, Berkeley was sued for bias against
women who had applied for admission to graduate schools there \cite{wiki}. 
The admission figures for the fall of 1973 showed that men applying were
more likely than women to be admitted, and the difference was so large
that it was unlikely to be due to statistical fluctuations. 
On the other hand, when examining the individual departments, it appeared that
no department was significantly biased against women. In fact,
most departments had a small but statistically significant bias
in favor of women \cite{Bic75}.	The explanation of the "paradox" is
relatively simple: the women tended to apply to competitive departments, 
with low rates of admission even among qualified applicants (such as in the
English Department), whereas men tended to apply to
less-competitive departments with high rates of admission among
the qualified applicants (such as in engineering and chemistry), and
overall this results in a smaller fraction of women admitted in total.


\section*{References}
\begin{thebibliography}{99}
\bibitem{Sim51} E. H. Simpson, J. Roy. Stat. Soc. B
{\bf 13}, 238 (1951).
\bibitem{Bly72} C. R. Blyth J. Am. Stat.  Ass. {\bf 67}, 364
(1972).
\bibitem{Bic75} P. J. Bickel, E. A. Hammel, J. W. O'Connell, 
Science {\bf 187}, 398 (1975).
\bibitem{Mes81} D. M. Messick, J. P. van de Geer, 
Psychol. Bull. {\bf 90}, 582 (1981). 
\bibitem{YSs1} S. Ramanana-Rahary, M. Zitt, R. Rousseau, 
Scientometrics {\bf 79}, 311 (2009).
\bibitem{YSm1} J. S. Chuang, O. Rivoire, S. Leibler, Science {\bf 323},
272 (2009).
\bibitem{YSm2} S. G. Baker, B. S. Kramer, J Women Health Gen-B {\bf
10}, 867 (2001).
\bibitem{Lip06} S. Lipovetsky, W. M. Conklin, Eur. J. Op. Res. {\bf 172}, 334
(2006).
\bibitem{Her11} M. A. Hernan, D. Clayton, N. Keiding,
Int. J. Epidem. {\bf 40}, 780 (2011).
\bibitem{Ald95} J. Aldrich, Stat. Sci. {\bf 10}, 364 (1995).
\bibitem{Cox03} D. R. Cox, W. Wermuth, J. Roy. Stat. Soc. B {\bf 65},
937 (2003).
\bibitem{Par00}
J. M. R. Parrondo, G. P. Harmer, D. Abbott, Phys. Rev. Lett. {\bf 85}, 
5226 (2000).
\bibitem{Buc02} J. Buceta, K. Lindenberg, J. M. R. Parrondo, 
Fluctuation Noise Lett. {\bf 2}, L21 (2002).
\bibitem{Kay03} R. J. Kay, N. F. Johnson, Phys. Rev. E {\bf 67}, 
056128 (2003).
\bibitem{Ast05} 
R. D. Astumian, M. Bier, Phys. Rev. Lett. {\bf 72}, 1766 (1994); 
R. D. Astumian, Am. J. Phys. {\bf 73}, 178 (2005). 
\bibitem{Gud88} S. P. Gudder, {\em Quantum probability}, (Academic
Press, Boston, 1988).
\bibitem{Coh86} J. E. Cohen, Am. Stat. {\bf 40}, 32 (1986).
\bibitem{Had97} P. Hadjicostas, Lin. Alg. Appl. {\bf 264}, 475 (1997).
\bibitem{Gal90} S. Galam, J. Stat. Phys. {\bf 61}, 943 (1990).
\bibitem{Alm05} J.  Almeida, D. Peralta-Salas, M. Romera, Physica D 
{\bf 200}, 124 (2005).
\bibitem{mcd09} M. D.  McDonnell, D. Abbot, Proc. Roy. Soc. A {\bf 465},
3309 (2009).
\bibitem{wiki} See more details and real figures at
{http://en.wikipedia.org/wiki/Simpson's\_paradox}
\end{thebibliography}
\end{document}